\documentclass[10pt,letterpaper]{article}
\usepackage{opex3}
\usepackage{color}
\usepackage{cite}
\usepackage{braket}
\begin{document}

\title{Second harmonic generation at 399 nm resonant on the $^{1}S_{0}-^{1}P_{1}$ transition of ytterbium using a periodically poled LiNbO$_{3}$ waveguide}

\author{Takumi Kobayashi,$^{1*}$ Daisuke Akamatsu,$^{1}$ Yoshiki Nishida,$^{2}$ Takehiko Tanabe,$^{1}$ Masami Yasuda,$^{1}$ Feng-Lei Hong,$^{3,1}$ and Kazumoto Hosaka$^{1}$}

\address{$^1$National Metrology Institute of Japan (NMIJ), National Institute of Advanced Industrial Science and Technology (AIST), 1-1-1 Umezono, Tsukuba, Ibaraki 305-8563, Japan\\
$^2$NTT Electronics Corporation, 6700-2 To, Naka-shi, Ibaraki 311-0122, Japan\\
$^3$Department of Physics, Graduate School of Engineering, Yokohama National University, 79-5 Tokiwadai, Hodogaya-ku, Yokohama 240-8501, Japan\\
}

\email{$^*$takumi-kobayashi@aist.go.jp} 



\begin{abstract}
We demonstrate a compact and robust method for generating a 399-nm light resonant on the  $^{1}S_{0}-^{1}P_{1}$ transition in ytterbium using a single-pass periodically poled LiNbO$_{3}$ waveguide for second harmonic generation (SHG). The obtained output power at 399 nm was 25 mW when a 798-nm fundamental power of 380 mW was coupled to the waveguide. We observed no degradation of the SHG power for 13 hours with a low power of 6 mW. The obtained SHG light has been used as a seed light for injection locking, which provides sufficient power for laser cooling ytterbium.
\end{abstract}



\section{Introduction}
\label{introductionsection}
Cold-atom experiments with ytterbium (Yb) atoms are of great interest in various research fields, including optical lattice clocks \cite{Kohno2009,Yasuda2012,Park2013,Hinkley2013}, quantum simulation \cite{Sugawa2011,Taie2012}, and quantum information processing \cite{Gorshkov2009,Daley2011}. Among these fields, optical lattice clocks have recently attracted considerable attention thanks to demonstrations of uncertainties at the 10$^{-18}$ level \cite{Bloom2014,Ushijima2015,Nicholson2015} and precise frequency ratio measurements of Yb and strontium (Sr) lattice clocks \cite{Akamatsu2014,Nemitz}. Such accurate clocks are expected to contribute to the redefinition of the SI second, the search for time variations of fundamental physical constants \cite{Blatt2008,Rosenband2008}, relativistic geodesy \cite{Chou2010}, and other tests of fundamental physics \cite{Krisher1996,Derevianko2014}. Some of these applications require robust transportable clocks. Several groups have therefore developed transportable optical lattice clocks \cite{Mura2013,Poli2014,Bongs2015}. 

In experiments with cold Yb atoms, high power light sources operating at 399 nm resonant on the $^{1}S_{0}-^{1}P_{1}$ transition are used to cool atoms to a millikelvin temperature. A compact and robust 399-nm light source is essential for cold Yb experiments including development of a transportable Yb optical lattice clock. Several groups have employed blue external cavity diode lasers (ECDL) emitting at 399 nm \cite{Kim2003,Kohno2008}. To obtain sufficient power for cooling Yb atoms, an injection locked 399-nm diode laser has usually been adopted to amplify the output power from a 399-nm ECDL \cite{Komori2003,Park2003,Hosoya2015}. Another approach to generating a 399-nm light has been to utilize the second harmonic generation (SHG) of a red laser operating at 798 nm, e.g., a 798-nm ECLD. A 798-nm diode laser is useful for constructing an external cavity laser, since the laser has good longitudinal and transverse modes and is widely available. Conventionally, a nonlinear optical crystal, e.g., a lithium triborate (LBO), a barium borate (BBO), a BiB$_{3}$O$_{6}$, or a periodically poled KTiOPO$_{4}$ crystal, has been placed in a power-buildup cavity to obtain a high SHG power \cite{Adams1992,Onoda2008,Pizzocaro2014,Ruseva2004,Wang2008,Wen2014,Han2014,Wen2016}. The output power at 399 nm using this method is typically higher than that of the blue diode laser system. However, this SHG scheme requires the locking of a high-finesse cavity, which increases the number of stabilization parts and thus compromises the compactness and robustness of a laser system. 

This paper presents the SHG of a 798-nm light using a single-pass periodically-poled LiNbO$_{3}$ (PPLN) waveguide. High conversion efficiency in a single-pass configuration can be obtained with a PPLN waveguide, since its waveguide structure allows a tightly confined beam to propagate a long distance in a PPLN crystal \cite{Jiang2009,Guo2013,Gege2016}. A PPLN waveguide is therefore attractive for developing a compact and robust laser system at 399 nm. Our previous reports regarding a single-pass PPLN waveguide include the generation of a 589-nm light with a power of 494 mW \cite{Nishikawa2009} and a 461-nm light with a power of 76 mW \cite{Akamatsu2011}. In the present work, an output power of 25 mW is generated at 399 nm when a fundamental power of 380 mW is coupled to the PPLN waveguide. The output power of a blue light near 400 nm is generally low, since the wavelength is close to the edge of ultraviolet absorption in a PPLN crystal. The fundamental power at 798 nm can easily be obtained using a robust and conventional setup consisting of a 798-nm ECDL and a tapered amplifier. By using the generated 399-nm light as a seed light for injection locking, a power of more than 200 mW can be obtained \cite{Hosoya2015}, which is sufficient for cooling Yb atoms.

\section{Experimental setup}
\begin{figure}[h]
\begin{center}
\includegraphics[width=10cm,bb=0 50 910 416]{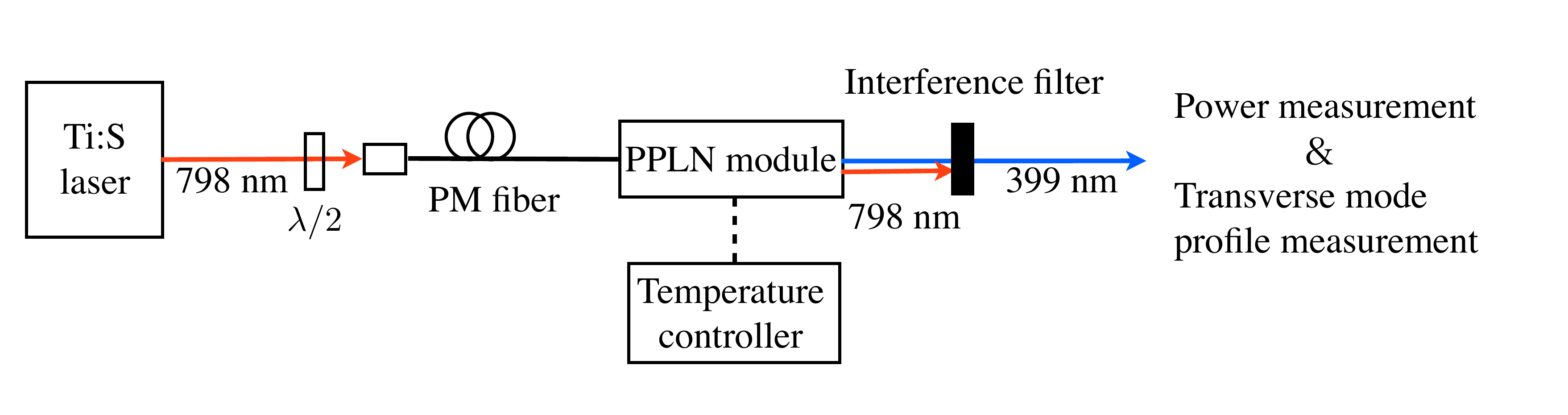}
\end{center}
\caption{Schematic diagram of the experimental setup. Ti:S: Titanium sapphire, $\lambda/2$: Half-wave plate, PM: Polarization-maintaining, PPLN: Periodically poled lithium niobate}
\label{experimentalsetup}
\end{figure}

Figure \ref{experimentalsetup} is a schematic diagram of the experimental setup. To evaluate the PPLN waveguide, we employed a titanium sapphire (Ti:S) laser (M Squared, SolsTiS) operating at 798 nm as a fundamental input light source. The 798-nm beam was allowed to pass through a half-wave plate and coupled into a polarization-maintaining (PM) fiber. The half-wave plate was used to adjust the polarization direction to maximize the coupling efficiency into the PM fiber, and hence the SHG light power. The total coupling efficiency of the fundamental light into the PPLN waveguide including the fiber coupling was about 40$\%$. The obtained SHG light was separated from the residual fundamental light using an interference filter. The PPLN waveguide was housed in a module package, a picture of which is shown in \cite{Nishikawa2009}. The temperature of the metal base of the PPLN waveguide was monitored with a thermistor and controlled at $<0.01$ $^{\circ}$C with a Peltier device to meet the phase matching condition.

The PPLN waveguide was manufactured by NTT Electronics Corp.  A 3-inch diameter 5-mol$\%$ ZnO-doped LiNbO$_{3}$ wafer of which material is highly resistant to photorefractive damage \cite{Asobe2001} and a 3-inch diameter LiTaO$_{3}$ wafer whose refractive index is lower than that of LiNbO$_{3}$ were used for the waveguide layer and substrate, respectively. Two wafers were brought into contact with each other without using any adhesives and then annealed at 500 $^{\circ}$C to achieve complete bonding. We used a conventional electrical poling method \cite{Myers1995} to fabricate the periodically poled structure. Since there is a resolution limit for a photoresist pattern of narrower than 2 $\mu$m using contact lithography, 3rd-order quasi-phase matching with periodically poling period of 7.625 $\mu$m was realized in advance on the ZnO-doped LiNbO$_{3}$. The waveguide layer thickness was reduced to 9.8 $\mu$m by successive lapping and polishing. We then fabricated a 11.6 $\mu$m-wide and 22 mm-long waveguide using a dicing saw \cite{Nishida2003}.

\section{Experimental results}
\label{experimentalresult}
\begin{figure}[b]
\begin{center}
\includegraphics[width=10cm,bb=100 20 910 716]{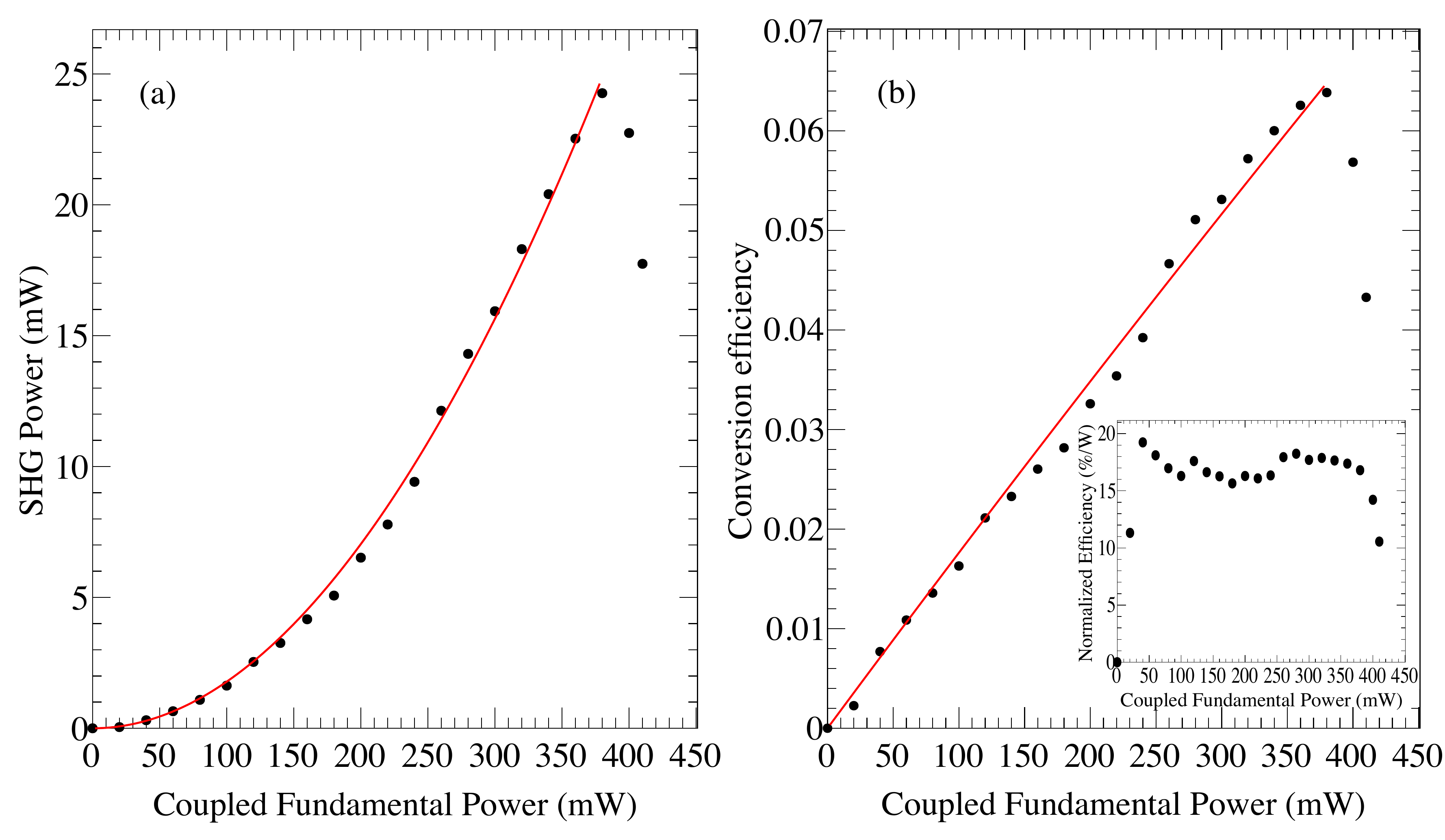}
\end{center}
\caption{SHG power (a) and conversion efficiency (b) as a function of the coupled fundamental power. The solid red curve indicates the best fit of a function based on the pump depletion model (see text). The inset in (b) shows the normalized conversion efficiency against the coupled fundamental power.}
\label{powerdependence}
\end{figure}
Figure \ref{powerdependence}(a) shows the measured SHG power as a function of the coupled fundamental power. The power value was corrected for the loss at the interference filter. A maximum power of 25 mW was obtained at 399 nm when the coupled fundamental power was 380 mW. The experimental data were fitted with the following formula based on the pump depletion model \cite{Parameswaran2002},
\begin{equation}
P^{2\omega}_{\mathrm{out}}=\eta P^{\omega}_{\mathrm{in}}=P^{\omega}_{\mathrm{in}}\tanh^{2}(\sqrt{\eta_{0}P^{\omega}_{\mathrm{in}}}L),
\end{equation}
where $P^{2\omega}_{\mathrm{out}}$ denotes the output power of the SHG light, $P^{\omega}_{\mathrm{in}}$ the input power of the fundamental light, $\eta$ the conversion efficiency, and $L$ the waveguide length. The normalized nonlinear efficiency $\eta_{0}$ is proportional to the square of the effective nonlinear coefficient $d_{\mathrm{eff}}$ (see Sect. \ref{discussion}). The fitting result for $P^{\omega}_{\mathrm{in}}\le 380$ mW is shown in Fig. \ref{powerdependence}(a) as a solid red line. The SHG power was found to decrease for $P^{\omega}_{\mathrm{in}}>380$ mW. We believe this is due to the absorption of the 399-nm light in the PPLN waveguide. This gives rise to a temperature gradient along the waveguide, which cannot be compensated for by the spatially uniform temperature control. Figure \ref{powerdependence}(b) shows the conversion efficiency $\eta$, with the fitting for $P^{\omega}_{\mathrm{in}}\le380$ mW. The maximum conversion efficiency was 6.4$\%$ when the coupled fundamental power was 380 mW. The inset in Fig. \ref{powerdependence}(b) shows the normalized conversion efficiency $\eta^{'}=P^{2\omega}_{\mathrm{out}}/(P^{\omega}_{\mathrm{in}})^{2}$. The normalized conversion efficiency was about 18$\%$/W.

\begin{figure}[h]
\begin{center}
\includegraphics[width=10cm,bb=100 20 910 516]{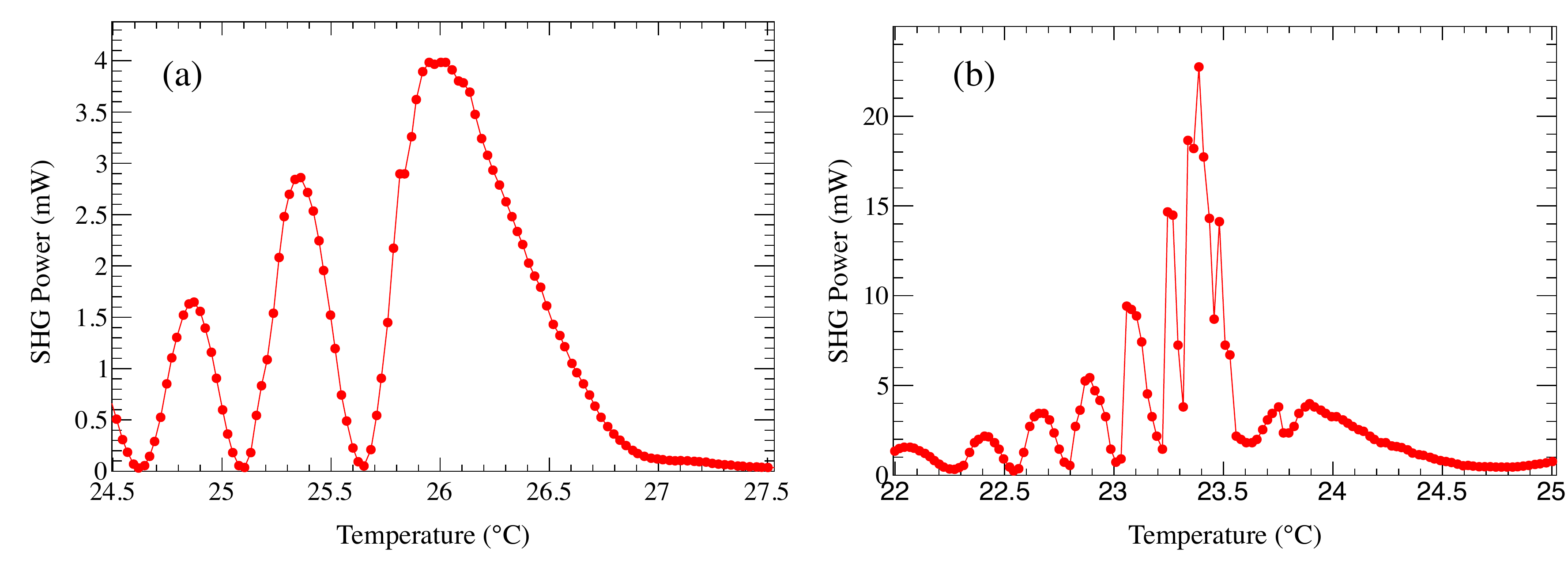}
\end{center}
\caption{Waveguide temperature dependence of the SHG power, where the coupled fundamental power is (a) 160 mW and (b) 400 mW.}
\label{temperaturedependence}
\end{figure}

We measured the waveguide temperature dependence of the SHG power. When the fundamental power was 160 mW, the temperature acceptance bandwidth of the phase matching was $\le0.5$ $^{\circ}$C as shown in Fig. \ref{temperaturedependence}(a). The acceptance became narrower ($<0.1$ $^{\circ}$C) for a high fundamental power of 400 mW as shown in Fig. \ref{temperaturedependence}(b). In this case, the waveguide temperature should be carefully controlled to obtain a stable SHG power. For example, the PPLN module package can be covered with a box to suppress temperature fluctuation \cite{Akamatsu2011}. The narrowing of the temperature acceptance is predicted by the pump depletion model \cite{Parameswaran2002} and has also been observed in our previous studies of a PPLN waveguide at 461 nm \cite{Akamatsu2011}. The temperature required to obtain the peak SHG power decreased by 2.6 $^{\circ}$C when the fundamental power was changed from 160 to 400 mW. This was because of the absorption of the 399-nm light inside the PPLN waveguide.

\begin{figure}[h]
\begin{center}
\includegraphics[width=8cm,bb=100 20 910 516]{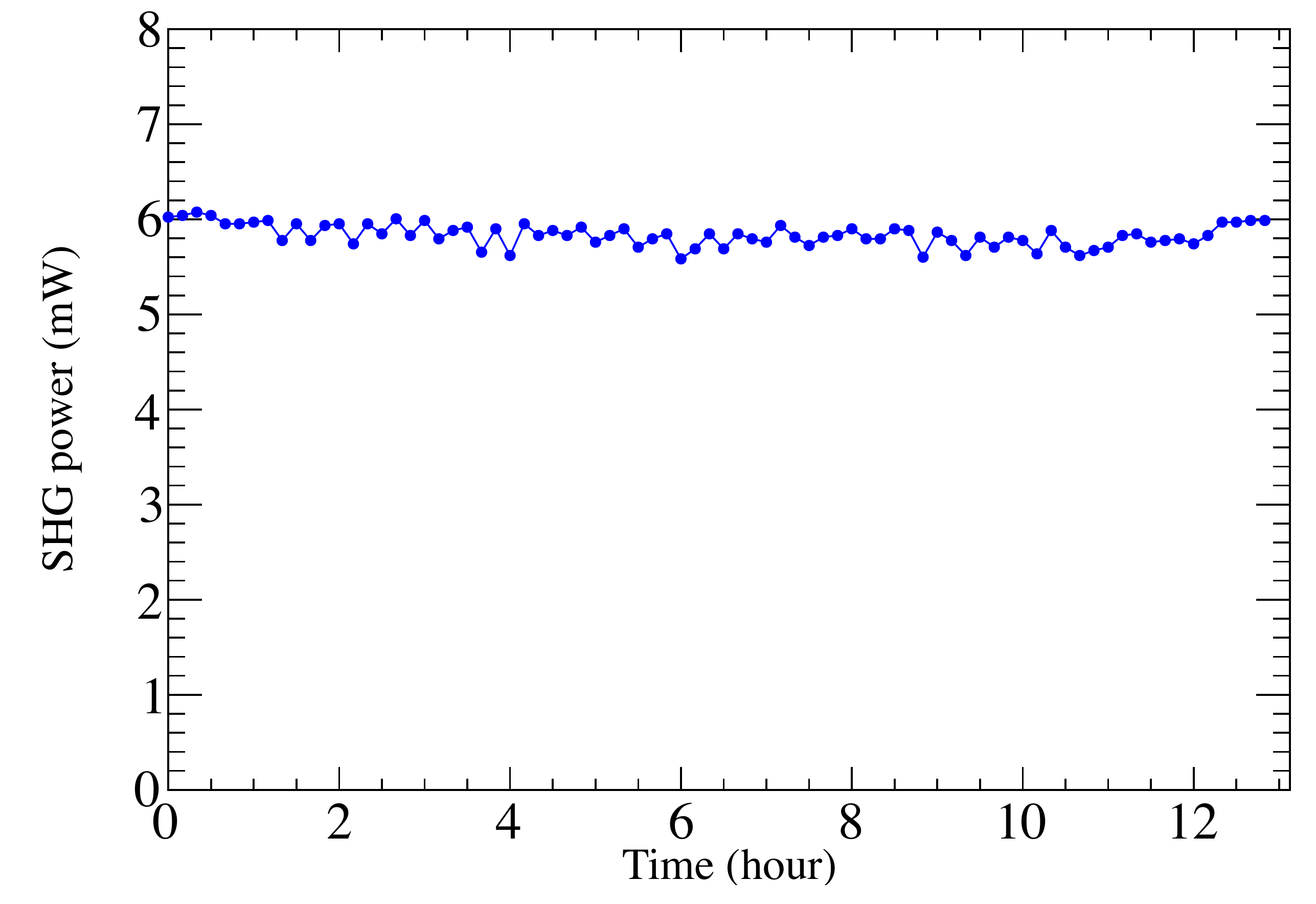}
\end{center}
\caption{Long-term stability of the SHG output power for a coupled fundamental power of 190 mW.}
\label{longterm}
\end{figure}

The long-term stability of the SHG power was investigated in the low power region where the temperature acceptance is relatively broad. Figure \ref{longterm} shows the SHG power as a function of time. The room temperature varied about $\pm1$ $^{\circ}$C during the measurement. No degradation of the SHG power was observed for 13 hours until we switched off the laser. The SHG power fluctuation was less than 8$\%$. This is considered to be due to the variation in the room temperature or the fluctuation of the coupling efficiency into the PM fiber. We also operated the PPLN module for several months and confirmed that the SHG power was stable day after day.

\begin{figure}[h]
\begin{center}
\includegraphics[width=11cm,bb=50 20 910 516]{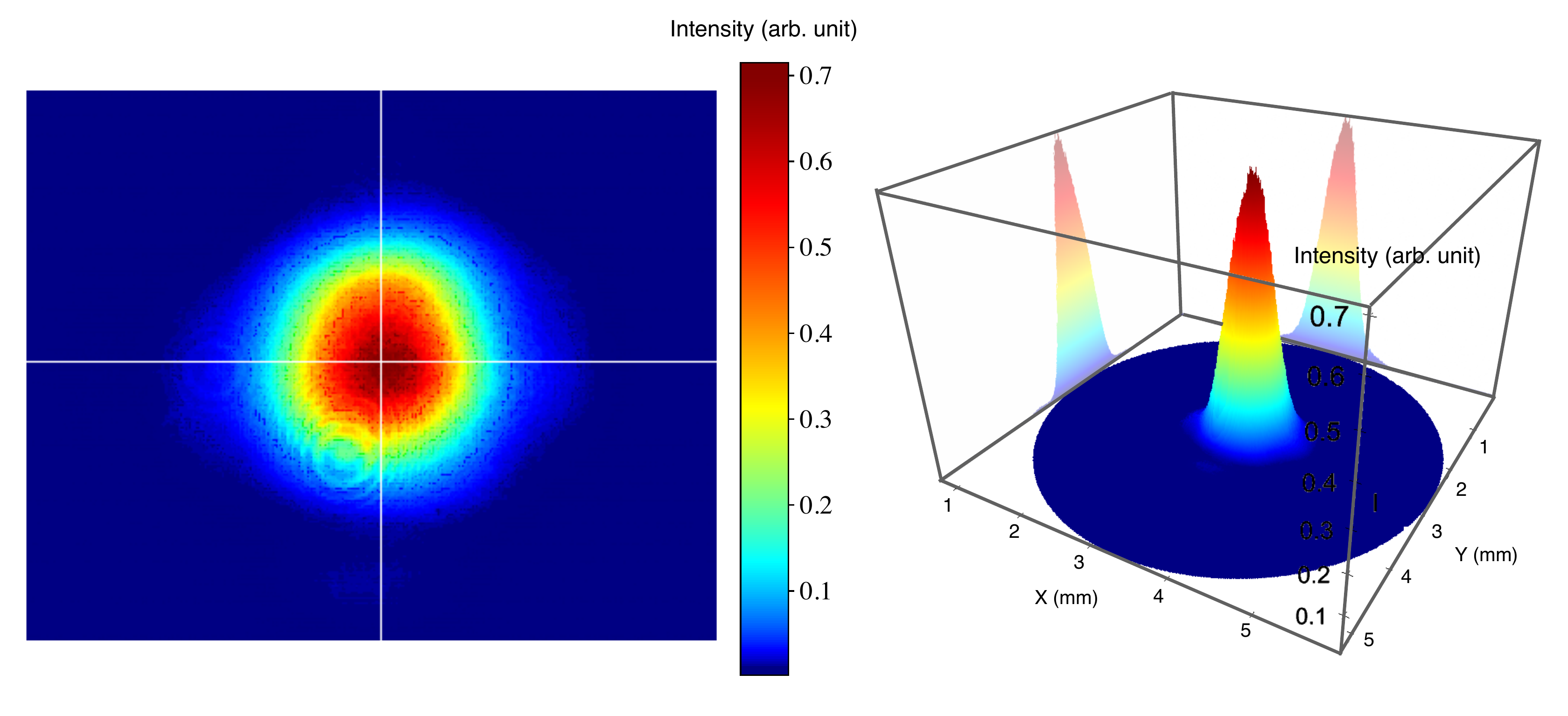}
\end{center}
\caption{Transverse mode profile of an SHG output beam.}
\label{transverse}
\end{figure}

Figure \ref{transverse} shows the transverse mode profile of the SHG output beam measured with a charge-coupled device camera. The profile was well-fitted by a Gaussian function. The ellipticity was 0.97. We confirmed that the coupling efficiency of the output beam into a single-mode fiber was about 70$\%$. 

\section{Discussion and conclusion}
\label{discussion}
The nonlinear coefficient $d_{\mathrm{eff}}$ of the PPLN waveguide based on 3rd-order quasi-phase matching was calculated to be about 3 pm/V from the fitting parameter $\eta_{0}$ (see Sect. \ref{experimentalresult}) and the relationship,
\begin{equation}
d_{\mathrm{eff}}=\frac{c}{\omega}\sqrt{\epsilon_{0} c n(\omega) n(2\omega)S}\sqrt{\eta_{0}},
\end{equation}
where $c$, $\epsilon_{0}$, $n(\omega)$, and $S$ are the speed of light, the vacuum permittivity, the refractive index of the waveguide for the fundamental frequency $\omega$, and the cross section of the waveguide, respectively \cite{Yariv}. In this calculation, the refractive indices $n(\omega)$ and $n(2\omega)$ were estimated using a Sellmeier equation with coefficients given in \cite{Jundt1997}. The calculated nonlinear coefficient of the present PPLN waveguide was found to be larger than that of a BBO (1.9 pm/V) \cite{Onoda2008} or an LBO crystal (0.75 pm/V) \cite{Pizzocaro2014} for the 399-nm SHG, but smaller than the typical coefficients of PPLN waveguides utilizing 1st-order quasi-phase matching ($\sim17$ pm/V) \cite{Shoji1997,Akamatsu2011}. It should be noted here that the nonlinear coefficient of 1st-order quasi-phase matching is theoretically three times larger than that of 3rd-order quasi-phase matching \cite{Yariv}.

The conversion efficiency obtained with the present PPLN waveguide was less than that achieved with a power-buildup cavity. For example, a 399-nm power of up to 1.0 W has been demonstrated from a fundamental power of 1.3 W, with a conversion efficiency of 80$\%$, using an LBO crystal in a cavity \cite{Pizzocaro2014}. The conversion efficiency in the present experiment was limited by (a) the relatively small nonlinear coefficient $d_{\mathrm{eff}}$ (see above) and (b) the reduction of the SHG power for fundamental powers $>380$ mW, which is caused by the absorption of the light in the PPLN waveguide (see Sect. \ref{experimentalresult}). The nonlinear coefficient can be increased by fabricating a poled structure with a short period (2.5 $\mu$m) needed for 1st-order quasi-phase matching. This fabrication is possible by using a high-voltage multi-pulse application method \cite{Mizuuchi2003}. The reduction or the saturation of the SHG power for a high fundamental power has already been observed using PPLN waveguides for relatively short wavelengths \cite{Bouchier2005,Akamatsu2011}. One possible way to reduce this effect is to fabricate a shorter waveguide. Such a waveguide provides a broader phase matching temperature acceptance \cite{Parameswaran2002}, and thus relaxes the requirement of the temperature uniformity along the waveguide. However, the reduction of the waveguide length shortens the interaction length for SHG. There must be an optimum waveguide length that provides a sufficient interaction length and a broad temperature acceptance bandwidth.

\begin{figure}[h]
\begin{center}
\includegraphics[width=12cm,bb=0 20 1010 716]{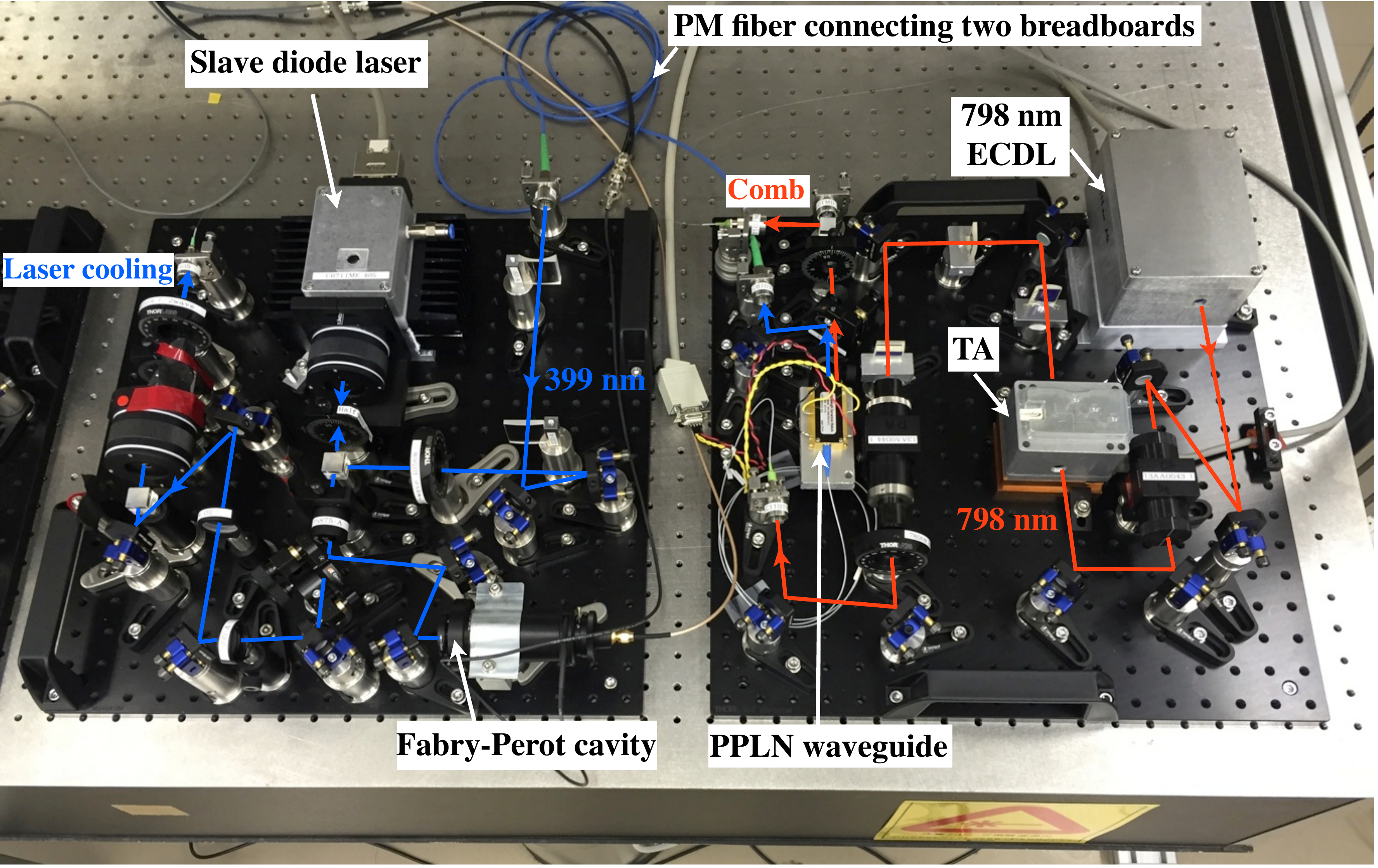}
\end{center}
\caption{Picture of our light source at 399 nm for laser cooling Yb atoms, consisting of a 798-nm external cavity diode laser (ECDL) with a tapered amplifier (TA) as a fundamental input laser, a periodically-poled LiNbO$_{3}$ (PPLN) waveguide, and an injection locked 399-nm slave diode laser. All the optical parts of the light source are arranged on two breadboards with areas of 45 cm $\times$ 45 cm. The seed light at 399 nm generated on the right breadboard is sent to the left breadboard via a polarization-maintaining (PM) fiber (blue jacket), and then injected into the slave laser. The residual fundamental 798-nm light transmitted from the PPLN waveguide is sent to an optical frequency comb through a PM fiber. The output beam at 399 nm is transported via a PM fiber for laser cooling.}
\label{picture}
\end{figure}

For the laser cooling of Yb atoms, the obtained SHG power from the PPLN waveguide is needed to be amplified by injection locking. In a previous experiment \cite{Hosoya2015}, a seed power of 5 mW at 399 nm was used to amply the laser power up to 220 mW. In the present experiment, we have already achieved a similar seed power (6 mW) with a good long-term stability. We have constructed a compact and robust light source at 399 nm that consists of an ECDL emitting at 798 nm with a tapered amplifier (TA) as a fundamental input laser, the present PPLN waveguide, and a slave diode laser for injection locking. In this laser system, active frequency stabilization is only needed for the 798-nm ECDL. One advantage of this laser system compared with an injection locked 399-nm slave diode laser using another blue ECDL is the utilization of a 798-nm diode laser which is useful for the reasons described in Sect. \ref{introductionsection}. Another advantage is that the 798-nm ECDL can easily be linked to an optical frequency comb based on a mode-locked erbium-doped fiber laser ($1000-2050$ nm) or a Ti:S laser ($500-1100$ nm). We have previously developed the fiber-based frequency combs for frequency stabilization of several light sources used in our Yb and Sr optical lattice clocks \cite{Yasuda2010,Akamatsu2012,Inaba2013}. The 798-nm ECDL can similarly be stabilized to our frequency comb by detecting a heterodyne beat note between the 798-nm light and an SHG comb component. Figure 6 shows a picture of our 399-nm system developed on two breadboards with areas of 45 cm $\times$ 45 cm that can be installed in a transportable 19-inch rack. We constructed a Littrow-type ECDL operating at 798 nm using a red diode laser (Sanyo DL-LS2075). The 798-nm light from the ECDL was amplified by a TA (Eagleyard Photonics EYP-TPA-0808) and then coupled into the PPLN waveguide. The generated SHG light was coupled into a PM fiber and injected into a salve blue diode laser (Nichia NDV4B16) after passing through lenses for optimizing transverse mode matching between the seed and slave beams. The temperature of the slave diode laser was set at $-4$ $^{\circ}$C to adjust its output wavelength to 399 nm at the operation current of $\sim$170 mA. To prevent condensation, the slave diode laser was housed in a box which was purged with nitrogen. A small portion of the slave beam was sent to a scanning Fabry-Perot cavity (Thorlabs SA200-3B) for monitoring the longitudinal mode. The details of the other experimental parameters are described in \cite{Hosoya2015}. Our laser system generated an output power of 200 mW at 399 nm when a seed power of 2 mW was injected into the slave laser. We successfully carried out laser cooling of Yb atoms using our laser system. 

In conclusion, we have demonstrated a compact and reliable method for generating a 399-nm light with a good transverse mode profile using a PPLN waveguide for SHG. A maximum power of 25 mW was obtained at 399 nm when the coupled fundamental power was 380 mW. At a relatively low SHG power of 6 mW, no degradation was observed in the SHG power for 13 hours. The obtained SHG light has been used as a seed light for injection locking, which generates sufficient power for laser cooling Yb atoms. The PPLN waveguide will be attractive for various experiments with cold Yb atoms including a transportable Yb optical lattice clock.

\section*{Acknowledgments}
We thank T. Hosoya for providing us information on injection locking. We are indebted to S. Okubo and M. Takamoto for helpful discussions regarding the simplification of optical systems. During the preparation of this article, we learned from S. Uetake
that he has performed similar experiments in Y. Takahashi's group in Kyoto University.  However, no related results
were either published in journals or presented at conferences.  We are grateful to S. Uetake for useful discussions. Part of this work was supported by Japan Society for the Promotion of Science (JSPS) KAKENHI Grant Number 15K05238, 24740281, and 25800231.
\end{document}